# Interstellar sulfur isotopes and stellar oxygen burning *

**Y.-N. Chin**[1], **C. Henkel**[2], **J.B. Whiteoak**[3], **N. Langer**[4], **and E.B. Churchwell**[5]

[1] Radioastronomisches Institut der Universität Bonn, Auf dem Hügel 71, D-53121 Bonn, Germany
[2] Max-Planck-Institut für Radioastronomie, Auf dem Hügel 69, D-53121 Bonn, Germany
[3] Australia Telescope National Facility, CSIRO Radiophysics Labs., P.O. Box 76, Epping, NSW 2121, Australia
[4] Max-Planck-Institut für Astrophysik, Karl-Schwarzschild-Str. 1, D-85740 Garching, Germany
[5] University of Wisconsin, Astronomy Department, 475 N. Charter St., Madison, WI 53706, USA

**Abstract.** A $^{12}C^{32}S$, $^{13}C^{32}S$, $^{12}C^{34}S$, and $^{12}C^{33}S$ $J$=2–1 line survey has been made to study interstellar $^{32}S/^{34}S$ and $^{34}S/^{33}S$ ratios from the galactic disk. The four CS isotopomers were detected in 20 star forming regions with galactocentric distances between 3 and 9 kpc. From a comparison of line velocities, the $C^{33}S$ $J$=2–1 rest frequency is $\sim$ 250 kHz below the value given in the Lovas (1992) catalog. Taking $^{12}C/^{13}C$ ratios from Wilson & Rood (1994) and assuming equal $^{12}C^{32}S$ and $^{13}C^{32}S$ excitation temperatures and beam filling factors, $^{12}C^{32}S$ opacities are in the range 3 to 15; average $^{32}S/^{34}S$ and $^{34}S/^{33}S$ isotope ratios are 24.4 $\pm$ 5.0 and 6.27 $\pm$ 1.01, respectively. While no systematic variation in the $^{34}S/^{33}S$ isotope ratio is found, the $^{32}S/^{34}S$ ratio increases with galactocentric distance when accounting for the $^{12}C/^{13}C$ gradient of the galactic disk. A fit to the unweighted data yields $^{32}S/^{34}S = 3.3 \pm 0.5(D_{GC}/\mathrm{kpc}) + 4.1 \pm 3.1$ with a correlation coefficient of 0.84. Since the interstellar sulfur (S) isotopes are synthesized by oxygen burning in massive stars, consequences for nucleosynthesis and models of chemical evolution are briefly discussed.

**Key words:** Nuclear reactions, nucleosynthesis, abundances – ISM: abundances – ISM: molecules

## 1. Introduction

All elements heavier than hydrogen are believed to be synthesized by nuclear processes. The standard Big Bang model can explain the bulk of the elemental abundances up to $^{7}Li$, while heavier nuclei must be synthesized in stars (e.g., Wilson & Rood 1994). Thus accurate measurements of elementary and isotopic abundances are needed to gain insight into the physical processes in the interior of stars.

While information on stellar and interstellar abundances can be obtained from optical spectroscopy, such studies of the atomic gas phase are mainly confined to the solar neighborhood and do not allow to discriminate between various isotopic constituents. This can only be done by observing molecular isotopomers. Such studies, carried out at cm- and mm-wavelengths, have the additional advantage that they are probing optically obscured regions across the entire Galaxy, thus providing information on its 'chemical' history. While the isotopic abundances of C, N, and O provide detailed information on CNO and helium burning (e.g., Wilson & Matteucci 1992; Henkel & Mauersberger 1993; Henkel et al. 1994b; Wilson & Rood 1994; Kahane 1995), sulfur (S) is a useful tool to study the late evolutionary stages of massive stars.

Among the sulfur bearing molecules, CS (carbon monosulfide) is the most suitable species: it is ubiquitous in the interstellar medium (ISM), its main isotopic lines are strong, and its lowest rotational transitions are observed at frequencies not requiring extremely favorable weather conditions. The three most prominent sulfur isotopes, $^{32}S$, $^{34}S$, and $^{33}S$ have solar system abundance ratios of 22.5 : 1 : 0.18 (e.g., Bogey et al. 1981; Kahane et al. 1988). These abundance ratios were also found in the late-type giant IRC+10216 (Wannier 1980; Kahane et al. 1988). The $^{32}S/^{34}S$ ratio was already measured with a beam size of 130″ toward star forming regions of the galactic disk and center (Frerking et al. 1980). During the last fifteen years, larger telescopes and more sensitive receivers have led to a higher quality of observational data. Also, interstellar $^{12}C/^{13}C$ ratios which are needed to obtain reliable $^{32}S/^{34}S$ values (see Sect. 4) are now much better known. Although $C^{33}S$ was detected by Sutton et al. (1985) and Cummins et al. (1986), no systematic study of interstellar $^{34}S/^{33}S$ ratios has yet been reported for the ISM. We thus observed the $J$=2–1 lines of $^{12}C^{32}S$ (here-

*Send offprint requests to*: Y.-N. Chin

* Based on observations with the Swedish-ESO Submillimeter Telescope (SEST) at the European Southern Observatory (ESO), La Silla, Chile

$C^{34}S$), and $^{12}C^{33}S$ (hereafter $C^{33}S$) toward a number of star forming regions. The $^{13}CS$, $C^{34}S$, and $C^{33}S$ lines allow to measure the relevant sulfur isotope column densities, while $^{12}CS$ and $^{13}CS$ are used to estimate the opacity of the $^{12}CS$ $J$=2–1 line.

## 2. Observations

The measurements were made in September 1993 and May 1994 using the SEST 15-m telescope at La Silla, Chile. At the frequencies of the $J$=2–1 transitions of $^{12}CS$, $^{13}CS$, $C^{34}S$, and $C^{33}S$ (97980.986, 92494.299, 96412.982, and 97172.086 MHz, respectively; Lovas 1992) the beamwidth was $\sim 53''$. We employed a Schottky receiver with system temperatures of order 500 K on a main beam brightness temperature scale ($T_{\rm MB}$). The backend was an acousto-optical spectrometer with 2000 contiguous channels and a channel separation of 43 kHz (0.13 – 0.14 km s$^{-1}$). All measurements were carried out in a dual beam switching mode (switching frequency 6 Hz) with a beam throw of $11'40''$ in azimuth. Integration times varied from 4 minutes for $^{12}CS$ to 40 minutes for $C^{33}S$.

The pointing accuracy, obtained from measurements of SiO maser sources, was better than $10''$. Using the software package CLASS for data reduction, all spectra were converted to a $T_{\rm MB}$ scale, adopting a main beam efficiency of $\eta_{\rm mb}$=0.74. Within a single observing period, line temperatures could be reproduced to 10%. Comparing $C^{34}S$ data from Sept. 1993 and May 1994, the $T_{\rm MB}$ scales are identical within 15%. $C^{34}S$ and $C^{33}S$ spectra from Orion-KL, taken in January 1995, also show only small deviations (<10%) from the results displayed in Fig. 1 and Table 1. This is consistent with the known long term stability of the 3 mm Schottky receiver which allowed to reproduce line intensities from NGC 4945 over a period of several years (cf. Henkel et al. 1990, 1994a; Whiteoak et al. 1990; Bergmann et al. 1992; Israel 1992; Dahlem et al. 1993). Furthermore we obtained in May 25, 1994, $C^{18}O$ and $C^{17}O$ SEST spectra and in June 1, 1994, $C^{18}O$ and $C^{17}O$ 30-m IRAM spectra from M 17 SW ($\alpha_{1950}$ = 18h 17$^{\rm m}$ 30$^{\rm s}$, $\delta_{1950} = -16°12'58''$). The SEST $C^{18}O/C^{17}O$ line intensity ratios obtained in a frequency switching and a dual beam switching mode (see above) were 3.43 and 3.50, respectively. The 30-m IRAM maps consisted of 16 positions with $20''$ spacing centered on the SEST position; these data were obtained in a position switching mode with the reference at an offset $(\Delta\alpha, \Delta\delta)$= ($20'$, 0). Convolving the IRAM spectra to a full width to half power beam size of $48''$ (the resolution of the SEST data), we obtain $I(C^{18}O)/I(C^{17}O)$ = 3.47. For a given position, the error in the individual $I(C^{18}O)/I(C^{17}O)$ ratios due to noise is $\sim$ 10% at the 15-m SEST and 30-m IRAM telescopes. The good agreement between the SEST and IRAM data lends further confidence to the results of this study.

Wood & Churchwell (1989) have shown that the colors of known ultracompact (UC) HII regions, i.e. recently formed O and B stars still embedded in their natal molecular cloud, are narrowly confined in the far infrared (FIR) color-color plots. They further showed that there are few other objects which share the FIR colors of the UC HII regions. Applying the two color criterion, it is then possible to identify embedded O and B stars anywhere in the galactic disk. From a complete sample of 74 IRAS sources with $\delta < -30°$ and $S_{100\mu{\rm m}}$>3000 Jy, we selected 20 objects with $T_{\rm MB}(C^{34}S) > 0.85$ K for further $^{12}CS$, $^{13}CS$, and $C^{33}S$ measurements. Besides these IRAS sources, Orion-KL and two positions toward the north-eastern part of the prominent star forming region NGC 6334 (radio source F (Rodriguez et al. 1982) and FIR-I (Straw & Hyland 1989)) were also observed. The distances of the southern IRAS objects to the galactic center, $D_{\rm GC}$, have been estimated with the following model:

$$\frac{\Theta(D_{\rm GC})}{\Theta_\circ} = 1.0074 \left(\frac{D_{\rm GC}}{D_\odot}\right)^{0.0382} + 0.00698 \,. \qquad (1)$$

$\Theta(D_{\rm GC})$ is the rotational velocity of a source at galactocentric distance $D_{\rm GC}$; $\Theta_\circ$ = 220 km s$^{-1}$ (Brand 1986) and $D_\odot$ = 8.5 kpc is the galactocentric distance of the Sun.

## 4. Results

The measured CS profiles are shown in Fig. 1 after a first order baseline subtraction and a removal of the image sideband $^{12}CS$ feature in the $C^{33}S$ spectra. The weak features sometimes seen at the blueshifted and redshifted sides of the $C^{33}S$ profile could not be identified (cf. Lovas 1988). From gaussian fits to the line parameters of NGC 6334 A, the frequency offsets of these weak features relative to $C^{33}S$ are +3.4 ± 0.1 and −2.3 ± 0.1 MHz, respectively (for $\nu(C^{33}S)$, see Sect. 2). The corresponding image sideband frequencies are $\sim$ 97991 and 97997 MHz. The CS line parameters obtained from gaussian fits are summarized in Table 1.

There is no significant difference in the radial velocities between $^{12}CS$, $C^{34}S$, and $^{13}CS$ ($\Delta\, v_{\rm LSR}(^{13}CS, ^{12}CS)$ = 0.15 ± 0.32 km s$^{-1}$; $\Delta\, v_{\rm LSR}(^{13}CS, C^{34}S)$ = 0.00 ± 0.21 km s$^{-1}$). However, the $C^{33}S$ velocities are different. In all 20 sources, it is $C^{33}S$ which has the highest velocity ($\Delta\, v_{\rm LSR}(C^{33}S, ^{12}CS)$ = 0.88 ± 0.27 km s$^{-1}$; $\Delta\, v_{\rm LSR}(C^{33}S, C^{34}S)$ = 0.73 ± 0.29 km s$^{-1}$; $\Delta\, v_{\rm LSR}(C^{33}S, ^{13}CS)$ = 0.73 ± 0.29 km s$^{-1}$). From our Orion-KL data taken in January 1995, $\Delta\, v_{\rm LSR}(C^{33}S, C^{34}S)$ = 0.46 ± 0.07 km s$^{-1}$. This is consistent with the line parameters displayed in Table 1. We conclude that the $C^{33}S$ $J$=2–1 frequency measured by us is $\sim$ 250 kHz higher than the calculated frequency given in the Lovas (1992) catalog. According to Bogey et al. (1981), the central and strongest group of hyperfine components is at $\nu$ = (97171.840 ± 0.030) MHz. This is 246

**Fig. 1.** $^{12}$CS, $^{13}$CS, C$^{34}$S, and C$^{33}$S spectra from our sample of 20 southern sources

**Table 1.** Line parameters from unsmoothed $J$=2–1 spectra of the four measured CS isotopomers

| Object | $\alpha$ (1950) ($^{\mathrm{h\ m\ s}}$) | $\delta$ (1950) ($^{\circ\ \prime\ \prime\prime}$) | Molecule | r.m.s. (mK) | $\int T_{\mathrm{MB}}\,\mathrm{d}v$ (K km s$^{-1}$) | $T_{\mathrm{MB}}$ (K) | $v_{\mathrm{LSR}}$ (km s$^{-1}$) | $\Delta v_{1/2}$ (km s$^{-1}$) | Velocity Range (km s$^{-1}$) |
|---|---|---|---|---|---|---|---|---|---|
| Orion-KL | 05 32 46.7 | −05 24 24 | $^{12}$CS | 139 | 131. ±0.42 | 16.8 | 8.6 | 5.1 | (-30, 40) |
| | | | $^{13}$CS | 83 | 5.42 ±0.16 | 0.899 | 8.4 | 5.0 | ( -5, 20) |
| | | | C$^{34}$S | 83 | 11.6 ±0.15 | 2.25 | 8.7 | 4.4 | ( -5, 20) |
| | | | C$^{33}$S | 72 | 1.54 ±0.09 | 0.394 | 9.3 | 3.6 | ( 3, 15) |
| IRAS 12320 | 12 32 01.7 | −61 22 52 | $^{12}$CS | 103 | 18.1 ±0.15 | 4.13 | -43.0 | 3.9 | (-51,-35) |
| | | | $^{13}$CS | 57 | 1.13 ±0.06 | 0.392 | -43.1 | 2.8 | (-47,-39) |
| | | | C$^{34}$S | 42 | 3.30 ±0.05 | 0.946 | -43.0 | 3.2 | (-48,-38) |
| | | | C$^{33}$S | 31 | 0.583±0.032 | 0.177 | -42.3 | 3.1 | (-46,-38) |
| IRAS 12326 | 12 32 41.0 | −62 45 57 | $^{12}$CS | 145 | 43.5 ±0.29 | 7.88 | -39.5 | 4.9 | (-55,-25) |
| | | | $^{13}$CS | 61 | 3.32 ±0.10 | 0.675 | -39.6 | 4.6 | (-50,-30) |
| | | | C$^{34}$S | 64 | 7.66 ±0.10 | 1.53 | -39.4 | 4.6 | (-50,-30) |
| | | | C$^{33}$S | 48 | 1.75 ±0.06 | 0.294 | -38.6 | 4.5 | (-45,-32) |
| IRAS 15290 | 15 29 01.2 | −55 46 06 | $^{12}$CS | 185 | 29.4 ±0.32 | 3.87 | -88.5 | 6.5 | (-99,-76) |
| | | | $^{13}$CS | 66 | 2.11 ±0.09 | 0.443 | -88.6 | 4.3 | (-95,-82) |
| | | | C$^{34}$S | 93 | 4.82 ±0.13 | 0.883 | -88.5 | 4.8 | (-96,-81) |
| | | | C$^{33}$S | 36 | 0.624±0.047 | 0.116 | -87.6 | 4.4 | (-94,-81) |
| IRAS 15408 | 15 40 53.0 | −53 56 31 | $^{12}$CS | 198 | 58.2 ±0.36 | 7.27 | -39.1 | 7.0 | (-52,-27) |
| | | | $^{13}$CS | 67 | 2.77 ±0.11 | 0.458 | -38.9 | 5.1 | (-48,-30) |
| | | | C$^{34}$S | 91 | 6.66 ±0.14 | 0.920 | -38.9 | 6.3 | (-48,-30) |
| | | | C$^{33}$S | 53 | 0.751±0.075 | 0.166 | -38.5 | 4.8 | (-46,-31) |
| IRAS 15491 | 15 49 13.0 | −54 26 30 | $^{12}$CS | 173 | 52.2 ±0.29 | 7.71 | -47.0 | 6.3 | (-58,-36) |
| | | | $^{13}$CS | 72 | 5.41 ±0.10 | 0.965 | -47.0 | 5.4 | (-54,-40) |
| | | | C$^{34}$S | 83 | 11.4 ±0.11 | 1.94 | -46.8 | 5.6 | (-54,-40) |
| | | | C$^{33}$S | 46 | 1.67 ±0.06 | 0.284 | -46.1 | 5.6 | (-53,-40) |
| IRAS 15520 | 15 52 00.1 | −52 34 26 | $^{12}$CS | 175 | 57.7 ±0.32 | 8.66 | -42.6 | 5.1 | (-55,-30) |
| | | | $^{13}$CS | 57 | 8.26 ±0.10 | 1.39 | -41.6 | 5.2 | (-52,-31) |
| | | | C$^{34}$S | 71 | 15.4 ±0.12 | 2.53 | -41.8 | 5.3 | (-52,-31) |
| | | | C$^{33}$S | 92 | 2.40 ±0.12 | 0.481 | -41.4 | 3.8 | (-48,-35) |
| IRAS 15567 | 15 56 43.5 | −52 36 19 | $^{12}$CS | 179 | 43.0 ±0.37 | 5.84 | -107.7 | 5.6 | (-124,-91) |
| | | | $^{13}$CS | 64 | 2.43 ±0.10 | 0.434 | -107.7 | 4.6 | (-115,-99) |
| | | | C$^{34}$S | 98 | 5.58 ±0.13 | 1.12 | -107.9 | 4.3 | (-115,-101) |
| | | | C$^{33}$S | 35 | 0.918±0.051 | 0.151 | -107.3 | 5.1 | (-115,-99) |
| IRAS 16065 | 16 06 32.3 | −51 58 15 | $^{12}$CS | 175 | 40.7 ±0.31 | 5.74 | -63.1 | 6.2 | (-75,-51) |
| | | | $^{13}$CS | 69 | 4.70 ±0.11 | 0.835 | -62.7 | 4.7 | (-72,-53) |
| | | | C$^{34}$S | 83 | 8.99 ±0.13 | 1.70 | -62.8 | 4.7 | (-72,-53) |
| | | | C$^{33}$S | 46 | 1.57 ±0.07 | 0.242 | -62.0 | 5.7 | (-70,-54) |
| IRAS 16164 | 16 16 26.2 | −50 46 10 | $^{12}$CS | 171 | 42.7 ±0.31 | 5.18 | -56.7 | 7.7 | (-69,-44) |
| | | | $^{13}$CS | 67 | 4.93 ±0.11 | 0.616 | -56.7 | 7.2 | (-67,-46) |
| | | | C$^{34}$S | 96 | 9.72 ±0.15 | 1.41 | -56.5 | 6.7 | (-66,-47) |
| | | | C$^{33}$S | 44 | 1.99 ±0.08 | 0.175 | -55.7 | 9.7 | (-68,-43) |

| Object | $\alpha$ (1950) ($^{h\ m\ s}$) | $\delta$ (1950) ($^{\circ}\ '\ ''$) | Molecule | r.m.s. (mK) | $\int T_{MB}\,dv$ ($K\,km\,s^{-1}$) | $T_{MB}$ (K) | $v_{LSR}$ ($km\,s^{-1}$) | $\Delta v_{1/2}$ ($km\,s^{-1}$) | Velocity Range ($km\,s^{-1}$) |
|---|---|---|---|---|---|---|---|---|---|
| IRAS 16172 | 16 17 13.3 | −50 28 14 | $^{12}CS$ | 97 | 95.4 ±0.26 | 11.1 | -51.9 | 7.3 | (-80,-25) |
| | | | $^{13}CS$ | 89 | 11.2 ±0.17 | 1.87 | -51.8 | 5.4 | (-65,-40) |
| | | | $C^{34}S$ | 68 | 26.3 ±0.12 | 4.00 | -52.1 | 5.9 | (-65,-40) |
| | | | $C^{33}S$ | 91 | 4.82 ±0.17 | 0.578 | -51.2 | 5.9 | (-65,-40) |
| IRAS 16506 | 16 50 39.2 | −45 12 43 | $^{12}CS$ | 163 | 24.2 ±0.24 | 5.39 | -26.7 | 4.3 | (-35,-19) |
| | | | $^{13}CS$ | 68 | 1.97 ±0.09 | 0.566 | -26.9 | 3.5 | (-33,-21) |
| | | | $C^{34}S$ | 92 | 4.81 ±0.12 | 1.10 | -26.8 | 4.0 | (-33,-21) |
| | | | $C^{33}S$ | 44 | 0.766±0.062 | 0.147 | -25.6 | 4.5 | (-33,-18) |
| IRAS 16562 | 16 56 14.1 | −39 59 15 | $^{12}CS$ | 92 | 74.2 ±0.18 | 13.1 | -12.3 | 5.1 | (-25, 5) |
| | | | $^{13}CS$ | 83 | 5.48 ±0.13 | 1.36 | -12.3 | 3.7 | (-21, -4) |
| | | | $C^{34}S$ | 84 | 12.0 ±0.13 | 2.74 | -12.4 | 4.0 | (-21, -4) |
| | | | $C^{33}S$ | 90 | 1.88 ±0.11 | 0.441 | -11.6 | 4.1 | (-18, -6) |
| IRAS 17009 | 17 00 59.9 | −40 42 19 | $^{12}CS$ | 172 | 45.6 ±0.27 | 7.09 | -17.4 | 5.8 | (-26, -7) |
| | | | $^{13}CS$ | 81 | 3.09 ±0.12 | 0.643 | -16.8 | 4.6 | (-24, -9) |
| | | | $C^{34}S$ | 89 | 8.66 ±0.13 | 1.74 | -16.8 | 4.6 | (-24, -9) |
| | | | $C^{33}S$ | 55 | 1.35 ±0.08 | 0.201 | -16.0 | 6.1 | (-24, -9) |
| IRAS 17059 | 17 05 59.8 | −41 32 09 | $^{12}CS$ | 103 | 32.8 ±0.14 | 9.01 | -21.4 | 3.2 | (-29,-14) |
| | | | $^{13}CS$ | 78 | 3.78 ±0.10 | 1.23 | -21.5 | 2.7 | (-27,-16) |
| | | | $C^{34}S$ | 55 | 8.33 ±0.07 | 2.68 | -21.4 | 2.9 | (-27,-16) |
| | | | $C^{33}S$ | 36 | 1.59 ±0.04 | 0.476 | -20.8 | 2.9 | (-27,-16) |
| IRAS 17160 | 17 16 02.6 | −37 07 51 | $^{12}CS$ | 157 | 34.4 ±0.27 | 4.78 | -69.8 | 6.3 | (-80,-58) |
| | | | $^{13}CS$ | 70 | 3.03 ±0.10 | 0.573 | -69.7 | 4.5 | (-76,-62) |
| | | | $C^{34}S$ | 93 | 6.25 ±0.14 | 1.10 | -69.5 | 5.0 | (-77,-61) |
| | | | $C^{33}S$ | 44 | 0.985±0.051 | 0.206 | -68.9 | 4.6 | (-74,-64) |
| NGC 6334 A | 17 17 32.0 | −35 44 05 | $^{12}CS$ | 87 | 102. ±0.18 | 15.7 | -7.0 | 5.7 | (-22, 8) |
| | | | $^{13}CS$ | 56 | 11.3 ±0.09 | 2.32 | -6.6 | 4.5 | (-15, 2) |
| | | | $C^{34}S$ | 83 | 22.6 ±0.12 | 4.53 | -6.8 | 4.5 | (-15, 2) |
| | | | $C^{33}S$ | 41 | 3.95 ±0.05 | 0.802 | -6.1 | 4.6 | (-13, -2) |
| NGC 6334 B | 17 17 34.0 | −35 42 08 | $^{12}CS$ | 86 | 83.8 ±0.18 | 9.53 | -4.2 | 7.9 | (-20, 12) |
| | | | $^{13}CS$ | 64 | 8.47 ±0.10 | 1.76 | -4.2 | 4.2 | (-12, 4) |
| | | | $C^{34}S$ | 80 | 18.1 ±0.12 | 3.62 | -4.3 | 4.6 | (-12, 4) |
| | | | $C^{33}S$ | 40 | 2.71 ±0.05 | 0.729 | -3.7 | 3.5 | (-10, 2) |
| IRAS 17233 | 17 23 22.3 | −36 06 55 | $^{12}CS$ | 98 | 57.3 ±0.21 | 6.70 | -3.3 | 7.2 | (-21, 14) |
| | | | $^{13}CS$ | 94 | 7.60 ±0.13 | 1.37 | -2.8 | 5.0 | (-10, 4) |
| | | | $C^{34}S$ | 58 | 15.5 ±0.10 | 2.49 | -2.8 | 5.4 | (-15, 9) |
| | | | $C^{33}S$ | 42 | 2.33 ±0.06 | 0.397 | -2.2 | 5.3 | ( -9, 4) |
| IRAS 17257 | 17 25 53.7 | −36 37 52 | $^{12}CS$ | 101 | 72.8 ±0.16 | 10.5 | -11.0 | 6.5 | (-21, -1) |
| | | | $^{13}CS$ | 75 | 6.36 ±0.12 | 1.05 | -10.5 | 5.4 | (-20, -2) |
| | | | $C^{34}S$ | 59 | 13.8 ±0.09 | 2.22 | -10.7 | 5.8 | (-20, -2) |
| | | | $C^{33}S$ | 37 | 2.24 ±0.05 | 0.388 | -9.7 | 5.3 | (-17, -5) |

catalog and agrees well with our observational result.

The main isotopomer, $^{12}$CS, is known to be moderately optically thick (e.g., Frerking et al. 1980; Linke & Goldsmith 1980). Therefore, the $^{32}$S/$^{34}$S isotope ratio cannot be determined from the observed $I(^{12}\mathrm{CS})/I(\mathrm{C}^{34}\mathrm{S})$ ratio. With the assumption that C$^{34}$S and $^{13}$CS are optically thin, however, $^{32}$S/$^{34}$S can be estimated from $I(^{13}\mathrm{CS})/I(\mathrm{C}^{34}\mathrm{S})$, if the $^{12}$C/$^{13}$C ratio is known:

$$\frac{^{32}\mathrm{S}}{^{34}\mathrm{S}} \sim \frac{^{12}\mathrm{C}}{^{13}\mathrm{C}}\,\frac{I(^{13}\mathrm{C}^{32}\mathrm{S})}{I(^{12}\mathrm{C}^{34}\mathrm{S})}. \qquad (2)$$

Assuming equal excitation temperatures and beam filling factors for $^{12}$CS and $^{13}$CS, the $^{12}$CS $J$=2–1 peak opacity $\tau$ ($^{12}$CS) can be determined from

$$\frac{T_{\mathrm{MB}}(^{12}\mathrm{CS})}{T_{\mathrm{MB}}(^{13}\mathrm{CS})} \sim \frac{1-\mathrm{e}^{-\tau(^{12}\mathrm{CS})}}{1-\mathrm{e}^{-\tau(^{12}\mathrm{CS})/R}}, \quad R = \frac{^{12}\mathrm{C}}{^{13}\mathrm{C}}, \qquad (3)$$

$^{12}$C/$^{13}$C ratios have been obtained from $H_2CO$ (e.g., Henkel et al. 1982, 1985) and C$^{18}$O (Langer & Penzias 1990) throughout the galactic disk. In the ISM, the carbon isotope ratio is thus the best known among the various CNO isotope ratios. Wilson & Rood (1994) obtain from a fit to these data

$$\frac{^{12}\mathrm{C}}{^{13}\mathrm{C}} = (7.5 \pm 1.9)\,(D_{\mathrm{GC}}/\mathrm{kpc}) + (7.6 \pm 12.9). \qquad (4)$$

For Orion-KL, $^{12}$C/$^{13}$C ratios of 79 ± 7 and 67 ± 3 were determined from C$^{18}$O (Langer & Penzias 1990). Because of possible errors in a single measurement, fractionation (see Sect. 5.2), and selfconsistency, Eq. 4 was also applied to the Orion-KL data.

While the $^{32}$S/$^{34}$S isotope ratio cannot be determined from CS alone, the $^{34}$S/$^{33}$S isotope ratio is directly obtained from

$$\frac{^{34}\mathrm{S}}{^{33}\mathrm{S}} \sim \frac{I(\mathrm{C}^{34}\mathrm{S})}{I(\mathrm{C}^{33}\mathrm{S})}. \qquad (5)$$

Eq. 5 requires that C$^{34}$S is optically thin (see Sect. 5.1).

In Table 2, galactocentric distances, $^{12}$CS $J$=2–1 peak opacities, and $^{12}$C/$^{13}$C, $^{32}$S/$^{34}$S, and $^{34}$S/$^{33}$S isotope ratios are derived from Eqs. 1–5 and Table 1 for our sample of sources. The average $^{32}$S/$^{34}$S and $^{34}$S/$^{33}$S ratios are with 24.4 ± 5.0 and 6.27 ± 1.01 (the errors are the standard deviations of the mean) consistent with the solar system values of 22.5 and 5.5.

## 5. Discussion

### *5.1. Photon trapping and line saturation*

To derive the quantities displayed in Table 2, unsaturated $^{13}$CS, C$^{34}$S, and C$^{33}$S $J$=2–1 lines as well as equal excitation temperatures ($T_{\mathrm{ex}}$) for all isotopomers have been assumed. In view of significant saturation and photon trapping in the main isotopic species, the validity of the two assumptions has to be examined in more detail.

temperatures ($T_{\mathrm{ex}}$) relative to the optically thin limit, can be estimated when calculating the statistical equilibrium populations of the various CS isotopes in the $J$=1 and 2 states. We have performed such calculations using the LVG (Large Velocity Gradient) approximation for a cloud of spherical geometry (e.g., Scoville & Solomon 1974). With collisional rates from Green & Chapman (1978), the populations for the 13 lowest rotational CS states have been determined. Table 3 provides some characteristic results for a kinetic temperature $T_{\mathrm{kin}}$= 30 K (cf. Padman et al. 1985), $n(\mathrm{H}_2)$= $10^5\,\mathrm{cm}^{-3}$ (cf. Mauersberger & Henkel 1989), and isotope ratios of 1 : 23, 1 : 60, and 1 : 126.5 relative to the main species. Note that saturation in C$^{34}$S will decrease the $I(\mathrm{C}^{34}\mathrm{S})/I(^{13}\mathrm{CS})$ and $I(\mathrm{C}^{34}\mathrm{S})/I(\mathrm{C}^{33}\mathrm{S})$ line intensity ratios below the corresponding abundance ratios. This effect is however significantly reduced by photon trapping. As exemplified by Table 3, the line intensity ratios should provide reliable isotope ratios up to $\tau$ (C$^{34}$S) ~ 1, where the error due to saturation becomes ~ 20% or less. In view of the $\tau_{\mathrm{max}}$ values given in Col. 5 of Table 2, this requirement appears to be fulfilled for most sources.

A direct test of the significance of C$^{34}$S line saturation to our data can be made by plotting the $I(^{13}\mathrm{CS})/I(\mathrm{C}^{34}\mathrm{S})$ and $I(\mathrm{C}^{33}\mathrm{S})/I(\mathrm{C}^{34}\mathrm{S})$ line intensity ratios as a function of $I(\mathrm{C}^{34}\mathrm{S})/I(^{12}\mathrm{CS})$. Sources with large values of $I(\mathrm{C}^{34}\mathrm{S})/I(^{12}\mathrm{CS})$ are more likely to be optically thick in C$^{34}$S, which should also enhance the $I(^{13}\mathrm{CS})/I(\mathrm{C}^{34}\mathrm{S})$ and $I(\mathrm{C}^{33}\mathrm{S})/I(\mathrm{C}^{34}\mathrm{S})$ ratios. As can be seen from Figs. 2a and b, no correlation is evident. This indicates that for most of the disk sources C$^{34}$S saturation is not significant, while a few optically thick sources cannot be excluded.

Another way to estimate the degree of saturation in the C$^{34}$S $J$=2–1 lines is a comparison of linewidths. We obtain for all 20 sources $\Delta$ ($\Delta v_{1/2}$ ($^{12}$CS,C$^{34}$S)) = 1.05 ± 0.74 km s$^{-1}$, $\Delta$ ($\Delta v_{1/2}$ (C$^{34}$S,$^{13}$CS)) = 0.19 ± 0.41 km s$^{-1}$, and $\Delta$ ($\Delta v_{1/2}$ (C$^{34}$S,C$^{33}$S)) = −0.04 ± 1.03 km s$^{-1}$. For the five spectra with the highest quality C$^{33}$S profiles (toward IRAS 15491, IRAS 16172, IRAS 17059, and NGC 6334 A and B), $\Delta$ ($\Delta v_{1/2}$(C$^{34}$S,C$^{33}$S)) = 0.16 ± 0.41 km s$^{-1}$. We conclude that while the saturation of $^{12}$CS is indicated by a higher linewidth ($\Delta v_{1/2}(^{12}\mathrm{CS}) > \Delta v_{1/2}$(C$^{34}$S, $^{13}$CS, C$^{33}$S) in 80% of our sources), there is no significant difference between the other isotopomers ($\Delta v_{1/2}$(C$^{34}$S) $>\Delta v_{1/2}$($^{13}$CS, C$^{33}$S) in only 40% of the sources). This result also suggests that C$^{34}$S line saturation is not important here.

### *5.2. Fractionation*

Observed isotope ratios cannot be used for an analysis of stellar nucleosynthesis and galactic 'chemical' evolution if the results are significantly modified by fractionation. The $^{12}$C/$^{13}$C isotope ratio is likely affected by the reaction

$$^{13}\mathrm{C}^{+} + {}^{12}\mathrm{CO} \longrightarrow {}^{12}\mathrm{C}^{+} + {}^{13}\mathrm{CO} + \Delta E_{35\,\mathrm{K}} \qquad (6)$$

**Table 2.** Sulfur isotope ratios and $^{12}$CS optical depths[a)]

| Source | $D_{GC}$ [b)] (kpc) | $^{12}C/^{13}C$ [c)] | $T_{MB}(^{12}CS)/T_{MB}(^{13}CS)$ [d)] | $\tau_{max}(^{12}CS)$ [e)] | $I(^{13}CS)/I(C^{34}S)$ | $^{32}S/^{34}S$ [f)] | $^{34}S/^{33}S$ [g)] |
|---|---|---|---|---|---|---|---|
| Orion-KL | 9.0 [h)] | 75±21 | 18.69±1.73 | 4.1 | 0.467±0.015 | 35±10 | 7.53±0.45 |
| IRAS 12320 | 7.0 [j)] | 60±19 | 10.53±1.55 | 6.0 | 0.342±0.019 | 21±7 | 5.66±0.32 |
| IRAS 12326 | 7.1 [j)] | 61±19 | 11.67±1.08 | 5.4 | 0.433±0.014 | 26±8 | 4.38±0.16 |
| IRAS 15290 | 5.0 [j)] | 45±16 | 8.74±1.37 | 5.4 | 0.438±0.022 | 20±7 | 7.72±0.62 |
| IRAS 15408 | 6.5 [j)] | 56±18 | 15.87±2.36 | 3.5 | 0.416±0.019 | 23±8 | 8.87±0.90 |
| IRAS 15491 | 6.1 [j)] | 53±17 | 7.99±0.62 | 7.1 | 0.475±0.010 | 25±8 | 6.83±0.25 |
| IRAS 15520 | 6.2 [j)] | 54±17 | 6.23±0.28 | 9.4 | 0.536±0.008 | 29±9 | 6.42±0.32 |
| IRAS 15567 | 4.3 [j)] | 40±15 | 13.46±2.03 | 2.9 | 0.435±0.021 | 17±7 | 6.08±0.37 |
| IRAS 16065 | 5.3 [j)] | 47±16 | 6.87±0.61 | 7.4 | 0.523±0.014 | 25±8 | 5.73±0.26 |
| IRAS 16164 | 5.4 [j)] | 48±16 | 8.41±0.96 | 6.1 | 0.507±0.014 | 24±8 | 4.88±0.21 |
| IRAS 16172 | 5.6 [j)] | 50±17 | 5.94±0.29 | 9.2 | 0.426±0.007 | 21±7 | 5.46±0.19 |
| IRAS 16506 | 6.2 [j)] | 54±17 | 9.52±1.18 | 6.0 | 0.410±0.021 | 22±7 | 6.28±0.53 |
| IRAS 16562 | 7.0 [j)] | 60±19 | 9.63±0.89 | 6.6 | 0.457±0.012 | 27±9 | 6.38±0.38 |
| IRAS 17009 | 6.5 [j)] | 56±18 | 11.03±1.41 | 5.3 | 0.357±0.015 | 20±6 | 6.41±0.39 |
| IRAS 17059 | 6.1 [j)] | 53±17 | 7.33±0.47 | 7.8 | 0.454±0.013 | 24±8 | 5.24±0.14 |
| IRAS 17160 | 2.9 [j)] | 29±14 | 8.34±1.06 | 3.6 | 0.485±0.019 | 14±7 | 6.35±0.36 |
| NGC 6334 A | 7.0 [i)] | 60±19 | 6.77±0.17 | 9.6 | 0.500±0.005 | 30±10 | 5.72±0.08 |
| NGC 6334 B | 7.0 [i)] | 60±19 | 5.41±0.20 | 12.3 | 0.468±0.006 | 28±9 | 6.68±0.13 |
| IRAS 17233 | 7.8 [j)] | 66±20 | 4.89±0.34 | 15.1 | 0.490±0.009 | 32±10 | 6.65±0.18 |
| IRAS 17257 | 6.3 [j)] | 55±18 | 10.00±0.72 | 5.8 | 0.461±0.009 | 25±8 | 6.16±0.14 |

a) The standard deviations for the $T_{MB}$, integrated line intensity, and isotope ratios (Cols. 4–8) do not include systematic calibration errors which are difficult to assess (for details, see Sect. 2).
b) The distance of the Sun to the galactic center, $D_{\odot}$, is assumed to be 8.5 kpc.
c) The $^{12}C/^{13}C$ ratio is estimated from the distance to the galactic center (see Eq. 4). The error is $\sigma = (1.9^2\, D_{GC}{}^2 + 12.9^2)^{1/2}$.
d) Peak temperature ratios; the uncertainties are derived from the $1\sigma$ noise level of the unsmoothed spectra (see Table 1). Note that the lines extend over several contiguous channels. The exceptionally large ratio toward Orion-KL is caused by the high velocity outflow.
e) Peak optial depths obtained from Eq. 3. No error is given, since more rotational CS transitions are needed for a successful simulation of the physical conditions (see Sect. 5.1). In the case of Orion-KL, the high velocity outflow seen in $^{12}$CS reduces the calculated optical depth substantially.
f) Calculated from Eqs. 2 and 4
g) Calculated with Eq. 5
h) see Langer & Penzias (1990)
i) See Gardner & Whiteoak (1979)
j) see Eq. 1

(Watson et al. 1976), which enhances $^{13}$CO in the more diffuse $C^+$ rich parts of molecular clouds. Eq. 4 is derived from both $C^{18}O$ and $H_2CO$ data, which are believed to bracket the actual carbon isotope ratio (e.g., Langer et al. 1984; Langer & Penzias 1990). For CS, the corresponding reaction

$$^{13}C^+ + {}^{12}CS \longrightarrow {}^{12}C^+ + {}^{13}CS + \Delta E_{25K}, \qquad (7)$$

may be less important. The bulk of the CS emission arises from the densest parts of the molecular clouds only (e.g., Snell et al. 1984; Mauersberger & Henkel 1989), where the relative abundance of $C^+$ may be smaller than in the more diffuse ISM traced by CO. Also, the kinetic temperatures to reach a given level of fractionation must be smaller for CS than for CO. Most important, however, CO can be fractionated and CO is more abundant than CS. The CO isotope exchange reaction (Eq. 6) thus enhances $^{13}$CO and, as it destroys $^{13}C^+$, it increases the $^{12}C^+/^{13}C^+$ ratio above the local $^{12}C/^{13}C$ ratio. The effective enhancement of $^{12}C^+$ over $^{13}C^+$ implies, that molecules forming from $C^+$ for which no isotope exchange is possible (e.g. CH, $H_2CO$, HCN, $C_2H$), will show larger $^{12}C/^{13}C$ ratios than CO and the local ISM. Since CS can be fractionated (Eq. 7), it is an intermediate case and we find

$$^{12}CO/^{13}CO < {}^{12}CS/^{13}CS < H_2^{12}CO/H_2^{13}CO \qquad (8)$$

for the abundance ratios of CO, CS, and $H_2CO$. In view of the moderate differences in the $^{12}C/^{13}C$ ratios derived

**Table 3.** Computed CS $J$=2–1 line parameters for $T_{\rm kin}$= 30 K

| $n$(CS)/(d$v$/d$r$)$^{a)}$ (cm$^{-2}$/km s$^{-1}$) | $n$(H$_2$) (cm$^{-3}$) | $T_{\rm ex}$ (K) | $\tau^{b)}$ | $T_{\rm MB}$$^{b)}$ (K) |
|---|---|---|---|---|
| 3.08 $10^{14}$ | $10^5$ | 18.1 | 4.52 | 14.70 |
| 1.34 $10^{13}$ | $10^5$ | 8.6 | 0.62 | 2.53 |
| 5.13 $10^{12}$ | $10^5$ | 7.7 | 0.28 | 1.10 |
| 2.43 $10^{12}$ | $10^5$ | 7.3 | 0.16 | 0.55 |

a) The ratios of the chosen CS column densities per km s$^{-1}$ are 60 : 2.6 : 1 : 0.47 (for $^{12}$CS, C$^{34}$S, $^{13}$CS, and C$^{33}$S, respectively).
b) At $T_{\rm kin}$= 30 K, the correlation between $T_{\rm MB}$ values and optical depths tends to be consistent with Eq. 3 for densities well in excess of $10^4$ cm$^{-3}$. Differences between results obtained in Eq. 3 and the LVG code are caused by variations in $T_{\rm ex}$ and by the use of Rayleigh-Jeans line temperatures in Eq. 3. The detailed radiative transfer (LVG) calculations allow to estimate excitation temperatures and thus provide those Planck temperatures which are actually measured.

from H$_2$CO and CO (e.g., Henkel et al. 1982; Langer & Penzias 1990), we conclude that the $^{12}$CS/$^{13}$CS abundance ratio should be close to the actual $^{12}$C/$^{13}$C ratio.

For isotope exchange reactions similar to Eqs. 6 and 7, but involving S$^+$, very low temperatures are required. The energies of the ground vibrational states differ between $^{12}$CS and C$^{34}$S (C$^{33}$S and C$^{34}$S) by 7.5 (4) K. This is not sufficient to allow for significant fractionation near sites of massive star formation.

### *5.3. The $^{32}$S/$^{34}$S gradient*

Fig. 3 displays the $^{32}$S/$^{34}$S and $^{34}$S/$^{33}$S isotope ratios as a function of galactocentric distance. According to Fig. 3a, there is strong evidence for the presence of a $^{32}$S/$^{34}$S gradient. A linear least squares fit to the unweighted data yields $^{32}$S/$^{34}$S= 3.3 $\pm$ 0.5 ($D_{\rm GC}$/kpc) + 4.1 $\pm$ 3.1 with a correlation coefficient of 0.84. However, the case is not as clear as it appears to be at first sight: Most of the sources are located within a small $D_{\rm GC}$ interval ranging from 5.5 to 7 kpc. The apparent increase in the $^{32}$S/$^{34}$S ratio with increasing $D_{\rm GC}$ is thus based on only half a dozen of sources outside that $D_{\rm GC}$ range and an independent confirmation of the gradient is clearly needed.

The $^{34}$S/$^{33}$S data do not show any hint for a systematic variation with galactocentric radius. We suspect that a significant part of the scatter in the ratios (almost a factor of 2) is due to the small signal-to-noise ratios in some C$^{33}$S spectra and blending with the nearby unidentified features (cf. Sect. 3). The image sideband $^{12}$CS profile in the C$^{33}$S spectra (not shown in Fig. 1; see Sect. 4) is not a problem.

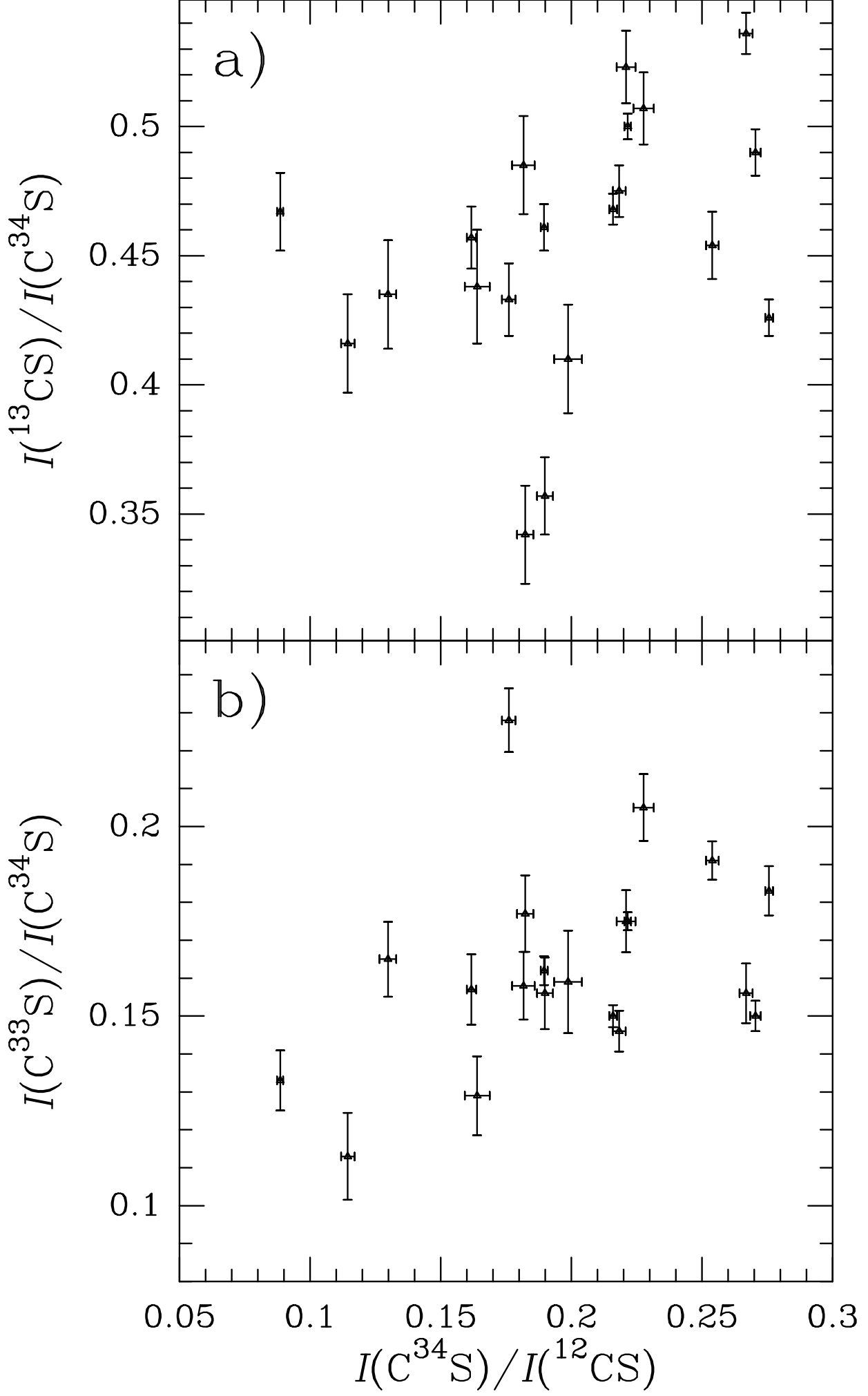

**Fig. 2. a)** $I$($^{13}$CS)/$I$(C$^{34}$S) and **b)** $I$(C$^{33}$S)/$I$(C$^{34}$S) as a function of $I$(C$^{34}$S)/$I$($^{12}$CS). In the case of saturated C$^{34}$S $J$=2–1 profiles, a positive correlation between the plotted intensity ratios is expected.

### *5.4. A comparison with other data*

Frerking et al. (1980) presented $^{12}$CS, $^{13}$CS, and C$^{34}$S $J$=2–1 data from 14 massive star forming regions. The two sources which are in common with our sample are Orion-KL and NGC 6334, where a position between the two observed by us was chosen. Their $I$($^{13}$CS)/$I$(C$^{34}$S) ratio toward Orion-KL, 0.36 $\pm$ 0.02, is $\sim$ 25% smaller than that given in Table 2; their ratio for NGC 6334 is with 0.52 $\pm$ 0.03 consistent with our values. Overall, Frerking et al. do not find strong evidence for C$^{34}$S saturation and their $I$($^{13}$CS)/$I$(C$^{34}$S) ratios from the galactic disk do not vary with galactocentric radius. These results are consistent with our data. The $^{32}$S/$^{34}$S gradient, not apparent in the $I$($^{13}$CS)/$I$(C$^{34}$S) line intensity ratios, is only seen when accounting for the interstellar $^{12}$C/$^{13}$C gradient across the galactic disk. Fifteen years ago, such a systematic varia-tion in the carbon ratio was not well established and was thus neglected by Frerking et al. (1980). The absence of a significant $I(^{13}CS)/I(C^{34}S)$ gradient with galactocentric radius and the agreement in these line intensity ratios ($0.42 \pm 0.16$ for the Frerking et al. and $0.45 \pm 0.10$ for our data; see Table 2) does not only provide some additional confirmation for the $^{32}S/^{34}S$ gradient proposed by us (see Sect. 5.3), but also shows that different angular resolutions (130″ versus 53″) do not qualitatively alter the properties of the line intensity ratios.

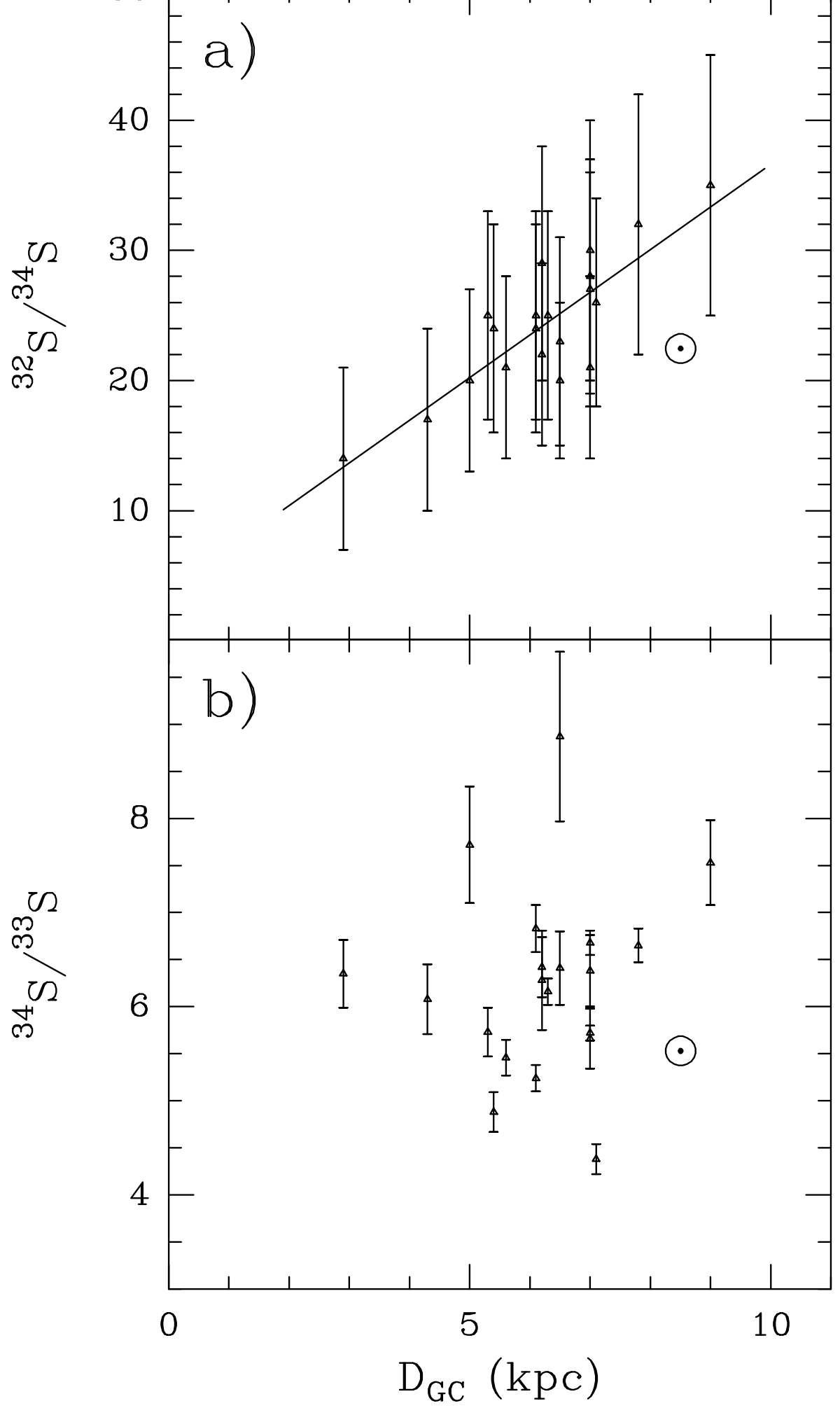

**Fig. 3. a)** $^{32}S/^{34}S$ and **b)** $^{34}S/^{33}S$ isotope ratios as a function of galactocentric radius. The solar system values at $D_{GC}$= 8.5 kpc are also given. The relatively large error bars in the $^{32}S/^{34}S$ data are caused by the uncertainty in the $^{12}C/^{13}C$ ratios (see Table 2). For the least square fit to the $^{32}S/^{34}S$ data displayed by a solid line, see Sect. 5.3.

As mentioned in Sect. 1, $C^{33}S$ was first observed by Sutton et al. (1985) and Cummins et al. (1986). For a 30″ region centered toward Orion-KL, Sutton et al. obtain from the $J$=5–4 transition $I(^{12}CS) : I(^{13}CS) : I(C^{34}S) : I(C^{33}S)$ = 291 : 16 : 33.5 : 11.0. While their $J$=5–4 corresponding $J$=2–1 line value, their $I(C^{34}S)/I(C^{33}S)$ ratio is only half as large as our value (see Table 2). Whether this discrepancy is caused by calibration uncertainties or other effects (e.g. ‘chemical’ inhomogeneities) remains to be seen. For the $J$=2–1 peak intensity ratios tward Orion-KL, the survey by Turner (1989) yields $T_{MB}(^{13}CS) : T_{MB}(C^{34}S) \sim 0.3 : 0.74$. This ratio is intermediate between the integrated line intensity ratios obtained by Frerking et al. (1980) and from our data (see Table 2).

### *5.5. Sulfur nucleosynthesis*

All the sulfur isotopes discussed in the present paper are believed to be of primary nature, synthesized during hydrostatic or explosive oxygen burning. These processes only take place in massive stars and Type Ia supernovae (SNe), with hydrostatic contributions dominating in the first, and solely explosive contributions dominating in the latter source (Thielemann & Arnett 1985, Weaver & Woosley 1993). The relative production ratios of $^{32}S$, $^{33}S$, and $^{34}S$ in massive stars appear to be rather independent of the initial stellar metallicity (Woosley & Weaver 1982). According to Thielemann et al. (1994) the present contribution of Type Ia SNe to the $^{32}S$ production is $\sim$ 40%. A comparison of Figs. 4 and 5 of Timmes et al. (1995) indicates a 20% contribution. While massive stars appear to slightly underproduce $^{33}S$ but overproduce $^{34}S$ compared to $^{32}S$ (Timmes et al. 1995; this holds equally well for SN Ib/c models which we do not distinguish from SN IIs in the following; cf. Woosley et al. 1995), the situation is just the opposite in Type Ia SNe (Thielemann et al. 1994).

These nucleosynthesis models do not favor galactic gradients of the sulfur isotopic ratios. Their primary nature, i.e. the independence of their production ratios from the initial stellar metal content, does not favor a relation of the sulfur isotopic ratios with the (controversial) galactic disk metallicity gradient (e.g., Matteucci & Francois 1989; Kaufer et al. 1994). A variation of the ratio of galactic Type II to Type Ia SN rates with galactocentric distance may also be problematic: This should lead to a significant variation of the $^{34}S/^{33}S$ ratio, which is not supported by our data (Fig. 3b), although the large scatter in the $^{34}S/^{33}S$ ratios does not exclude *small* (<50%) variations. A galactic $^{32}S/^{34}S$ gradient may thus imply that one of the basic assumptions in the generally accepted nucleosynthesis or chemical evolution models needs to be dropped. One possibility is that different kinds of Type Ia SNe, like Chandrasekhar mass and sub-Chandrasekhar mass models (cf. Woosley & Weaver 1995) yield different sulfur isotope production ratios, a question which still needs to be explored. If so, the data presented here might constrain their spatial distribution in the Milky Way. On the other hand, even the modelling of hydrostatic oxygen burning in massive stars may be generally oversimplified, as recently shown by the multi-dimensional calculations of Bazan &

sulfur production are still to be investigated.

In any case, the data presented here may turn out to provide an important test for the oxygen burning nucleosynthesis. This is another reason (cf. Sect. 5.3) why more observational support for the $^{32}S/^{34}S$ gradient is still needed. Sources of the northern hemisphere with well know distances from the Sun are suitable targets for a future study. Such observations would also allow to investigate (and to exclude) a Malmquist bias which *might* be caused by different spatial distributions of the various isotopomers (when analysing CNO isotope ratios, this effect was found not to be important; the effect is also not apparent when comparing CS data obtained with $130''$ and $53''$ beam sizes; see Sect. 5.4). Dark clouds, where ejecta from massive stars have not recently contaminated the ISM, may yield different ratios. Interesting information can also be obtained from the galactic center region. Accepting the gradient as shown in Fig. 3a with the solar system ratio well below the local interstellar value, we note that the results are similar to those obtained for the $^{14}N/^{15}N$ ratio (Dahmen et al. 1995).

## 6. Conclusions

Having studied the $J$=2–1 line profiles of $^{12}$CS, $^{13}$CS, C$^{34}$S, and C$^{33}$S in 20 prominent star forming regions, we obtain the following main results:

(1) From the CS data and previously determined $^{12}C/^{13}C$ ratios (Wilson & Rood 1994), we obtain a $^{32}S/^{34}S$ gradient of $3.3 \pm 0.5\,(D_{GC}/\mathrm{kpc})$ for the inner galactic disk out to the solar circle ($3\,\mathrm{kpc} \leq D_{GC} \leq 9\,\mathrm{kpc}$). Since only a few of our sources are located outside the range $5.5 \leq D_{GC} \leq 7\,\mathrm{kpc}$, this result has to be confirmed by additional measurements covering the less well sampled galactocentric distances and including northern sources to firmly exclude a possible Malmquist bias. A $^{32}S/^{34}S$ gradient is not expected in current theoretical schemes explaining sulfur nucleosynthesis in massive stars and Type Ia supernovae.

(2) Models of stellar nucleosynthesis predict that $^{33}$S enrichment is dominated by Type Ia supernovae, while a larger part of the $^{34}$S synthesis occurs in massive stars. In view of this difference, the absence of a *strong* galactic disk $^{34}S/^{33}S$ gradient is an interesting result.

(3) Assuming equal excitation temperatures and beam filling factors, peak $^{12}$CS $J$=2–1 opacities fall into the range 3–15.

(4) The calculated $J$=2–1 C$^{33}$S frequency given in the Lovas (1992) catalog appears to be too high. A correction of $\sim 250\,\mathrm{kHz}$ is suggested by the data.

(5) Additional measurements not only including northern sources, but also the galactic center region and dark clouds may add crucial information on sulfur isotope ratios and on oxygen burning in Type Ia supernovae and evolved massive stars.

F.-K. Thielemann, and A. Castets for contributing valuable comments and suggestions. Y.-N. Chin is a scholar of DAAD (German Academic Exchange Service, number 573 307 0023). N. Langer also thanks for support through DFG grant La 587/8-1.

## References

Bazan G., Arnett W.D., 1995, preprint
Bergman, P., Aalto, S., Black, J.H., Rydbeck, G., 1992, A&A 265, 403
Bogey, M., Demuynck, C., Destombes, J.L., 1981, Chem. Phys. Lett. 81, 256
Brand, J., 1986, Ph.D. Thesis, University of Leiden
Cummins, S.E., Linke, R.A., Thaddeus, P., 1986, ApJS 60, 819
Dahlem, M., Golla, G., Whiteoak, J.B., Wielebinski, R., Hüttemeister, S., Henkel, C., 1993, A&A 270, 29
Dahmen, G., Wilson, T.L., Matteucci, F., 1995, A&A, 295, 194
Frerking, M.A., Wilson, R.W., Linke, R.A., Wannier, P.G., 1980, ApJ 240, 65
Gardner, F.F., Whiteoak, J.B., 1979, MNRAS 188, 331
Green, S., Chapman, S., 1978, ApJS 37, 169
Henkel, C., Güsten, R., Gardner, F.F., 1985, A&A 143, 148
Henkel, C., Mauersberger, R., 1993, A&A 274, 730
Henkel, C., Whiteoak, J.B., Mauersberger, R., 1994a, A&A 284, 17
Henkel, C., Whiteoak, J.B., Nyman, L.-*Astron. Astrophys.*, Harju, J., 1990, A&A 230, L5
Henkel, C., Wilson, T.L., Bieging, J.H., 1982, A&A 109, 344
Henkel, C., Wilson, T.L., Langer, N., Chin, Y.-N., Mauersberger, R., 1994b, The Structure and Content of Molecular Clouds, eds. T.L. Wilson, K.J. Johnston, Springer Verlag (Lecture Notes in Physics 439), Springer-Verlag, Berlin, p72
Israel, F.P., 1992, A&A 265, 487
Kahane, C., 1995, Nuclei in the Cosmos, Proc. of the 3rd International Symposium on Nuclear Astrophysics, in press
Kahane, C., Gomez-Gonzalez, J., Cernicharo, J., Guélin, M., 1988, A&A 190, 167
Kaufer, A., Szeifert, Th., Krenzin, R., Baschek, B., Wolf, B., 1994, A&A 289, 740
Langer, W.D., Graedel, T.E., Frerking, M.A., Armentrout, P.B., 1984, ApJ 277, 581
Langer, W.D., Penzias, A.A., 1990, ApJ 357, 477
Linke, R.A., Goldsmith, P.F., 1980, ApJ 235, 437
Lovas, F.J., 1988, Frequency Catalog
Lovas, F.J., 1992, J. Phys. Chem. Ref. Data 21, 181
Matteucci, F., Francois, P., 1989, MNRAS 239, 885
Mauersberger, R., Henkel, C., 1989, A&A 223, 79
Padman, R., Scott, R.F., Vizard, D.R., Webster, A.S., 1985, MNRAS 214, 251
Rodriguez, L.F., Cantó, J., Moran, J.M., 1982, ApJ 255, 103
Scoville, N.Z., Solomon, P.M., 1974, ApJ 187, L67
Snell, R.L., Mundy, L.G., Goldsmith, P.F., Evans, N.J., Erickson, N.R., 1984, ApJ 276, 625
Straw, S.M., Hyland, A.R., 1989, ApJ 342, 876
Sutton, E.C., Blake, G.A., Masson, C.R., Phillips, T.G., 1985, ApJS 58, 341
Thielemann, F.-K., Arnett W.D., 1985, ApJ 295, 604

Proc. Les Houches Summer School, ed. S. Buldman et al., Elesevar Sci. Publishers B.V., Amsterdam, p. 629
Timmes F.X., Woosley S.E., Weaver T.A., 1995, ApJS, in press
Turner, B.E, 1989, ApJS 70, 539
Watson, W.D., Anichich, V.G., Huntress, W.T., 1976, ApJ 205, L165
Weaver, T.A., Woosley, S.E., 1993, Phys. Rep. 227, 65
Whiteoak, J.B., Dahlem, M., Wielebinski, R., Harnett, J.I., 1990, A&A 231, 25
Wilson, T.L., Matteucci, F., 1992, A&AR 4, 1
Wilson, T.L., Rood, R.T., 1994, ARA&A 32, 191
Wood, D.O.S., Churchwell, E.B., 1989, ApJ 340, 265
Woosley S.E., Weaver T.A., 1982, in *Nuclear Astrophysics*, ed. C.A. Barnes et al., Cambridge Univ. Press, p. 377
Woosley S.E., Weaver T.A., 1995, ApJ, in press
Woosley S.E., Langer N., Weaver T.A., 1995, ApJ, in press